\documentclass[12pt,twoside]{article}
\usepackage{fleqn,espcrc1}
\usepackage[final]{graphics}
\usepackage{amsfonts}

\title{Photon production by a quark-gluon plasma\thanks{Work done in
    collaboration with P.~Aurenche and H.~Zaraket.}}

\author{Fran\c cois Gelis\address[FG]{Brookhaven National Laboratory,
  Nuclear Theory, Bldg 510A, Upton, NY-11973, USA}}

\begin{document}
\maketitle

\begin{abstract}
  In this talk, I review the status of the calculation of the
  photon production rate in a hot quark-gluon plasma.  Particular
  emphasis is given to the discussion of the various length scales of
  the problem.
\end{abstract}

\section{Introduction - Model}
Photon or dilepton production by a hot plasma is expected to reflect
rather cleanly the state of the system. Indeed, in the collision of
two heavy nuclei, the typical size of the system is much smaller than
the mean free path of particles that feel only the electromagnetic
interactions, which enables photons and leptons to escape
freely. However, it has been realized recently that those rates
are non-perturbative quantities. In other words, even for an idealized
situation where the QCD coupling constant is small, the rate at
leading order receives contributions from an infinite set of diagrams.

To keep the model as simple as possible, I assume in this talk that
the temperature is much larger than all the quark masses ($T\gg m_{\rm
q}$), that the strong coupling constant at this temperature scale is
extremely small ($g\ll 1$), and that the system is in both thermal and
chemical equilibrium. Under these conditions, thermal field theory
seems to be the tool of choice to calculate the photon production
rate, which is obtained by calculating the imaginary part of the
retarded photon polarization tensor \cite{rate}:
\begin{equation}
{{dN^\gamma}\over{dtdV\,d\omega d^3{\mathbf q}}}\propto
 {\rm Im}\,\Pi^\mu_\mu{}_{\rm ret}(\omega,{\mathbf q})\; .
\end{equation}
In turn, the discontinuity of the retarded self-energy can be obtained
directly by using the cutting-rules of the retarded/advanced formalism
\cite{Gelis3}.  This formula automatically sums over all the possible
processes and interferences thereof at a given order in $g$ (and takes
care of the statistical weights).

\section{Lowest order}
At lowest non trivial order in the loop expansion for ${\rm
Im}\,\Pi^\mu_\mu{}_{\rm ret}$, the relevant processes for the
production of hard photons are the ones depicted on the following
figure: \setbox1=\hbox to
\textwidth{\hfill\resizebox*{!}{21mm}{\includegraphics{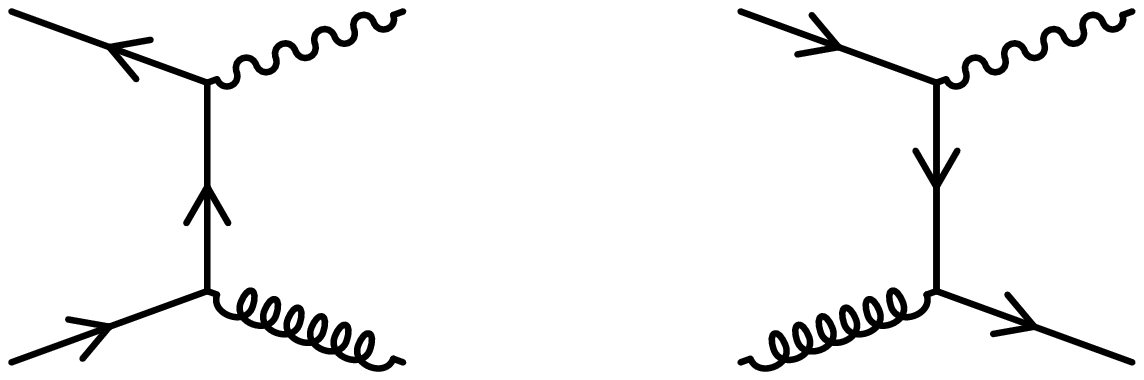}}\hfill}
\begin{displaymath}
\raise -11mm\box1
\end{displaymath}
Ignoring prefactors, their contribution to ${\rm
  Im}\,\Pi^\mu_\mu{}_{\rm ret}$ behave like \cite{1-loop}
\begin{equation}
{\rm Im}\,\Pi^\mu_\mu{}_{\rm ret}(\omega,{\mathbf q})\propto e^2 g^2 T^2 
\ln\Big({{\omega T}\over{m^2_{\rm th}}}\Big)\; ,
\end{equation}
where $m_{\rm th}\sim gT$ is a thermal mass of order $gT$. Such a
regulator is needed here, because the production rate of {\sl
massless} photons exhibit a logarithmic collinear singularity if
evaluated with massless quarks and gluons. The resummation of hard
thermal loops \cite{HTL} naturally provides this thermal mass.


\section{Bremsstrahlung-like processes}
One can note that a process like bremsstrahlung never appears in the
lowest order diagram. It shows up only in the next order (i.e. at
two-loop in the effective theory resumming the HTLs), together with a
process that differs from bremsstrahlung by crossing symmetry:
\setbox1=\hbox to
\textwidth{\hfill\resizebox*{!}{17mm}{\includegraphics{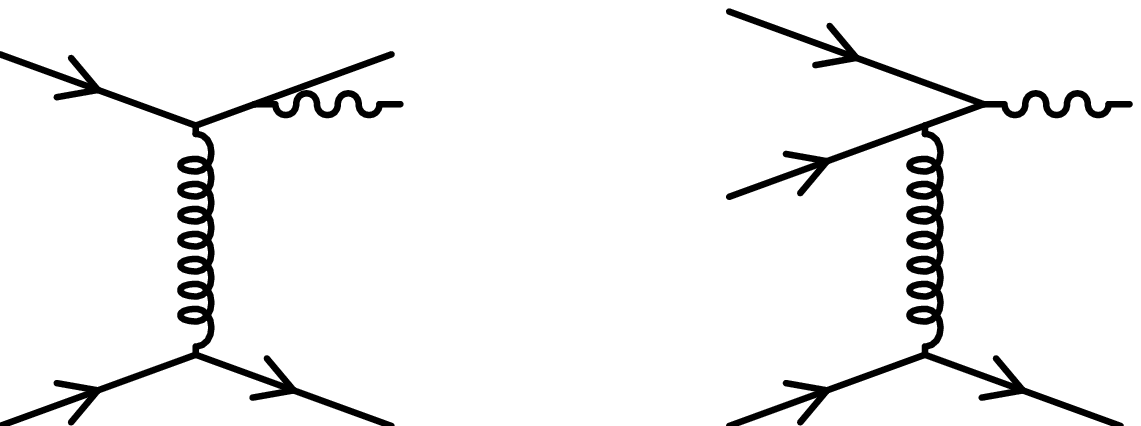}}\hfill}
\begin{displaymath}
\raise -11mm\box1
\end{displaymath}
The contribution of these processes to ${\rm Im}\,\Pi^\mu_\mu{}_{\rm
  ret}$ can be written in the following form
\cite{2-loop}:
\begin{equation}
{\rm Im}\,\Pi^\mu_\mu{}_{\rm ret}(\omega,{\mathbf q})\propto
e^2 g^4 {{T^2}\over{m^2_{\rm th}}}
\Big[{{T^3}\over{\omega}}\oplus T\omega\Big]\; ,
\end{equation}
where the first term dominates the low energy part of the spectrum,
while the second term dominates its high energy part. Naively, one
would expect those processes to come with four powers of $g$. However,
this is altered by very strong collinear singularities, that bring the
factor\footnote{This collinear factor decreases very fast if the
photon has a non-zero invariant mass. When the invariant mass is
maximal for a given energy, this factor is of order $1$
\cite{AurenGKZ2}.}  $T^2/m^2_{\rm th}\sim g^{-2}$ after regularization
by a thermal mass.  Therefore, it turns out that these contributions
are as large as the 1-loop ones (or even larger for soft photons).
The natural question one should ask after observing this collinear
enhancement at two loops is whether it happens also in higher order
diagrams, and whether the perturbative expansion can be kept under
control.

\section{Photon formation time and other scales}
The collinear singularities in photon production diagrams are
controlled by the virtuality of the quarks. It is particularly
instructive to calculate the virtuality of an off-shell quark of
momentum $R\equiv P+Q$ splitting into a photon of momentum $Q$ (and
invariant mass $Q^2$) and an on-shell quark of momentum $P$
\cite{Gelis11}:
\begin{equation}
R^2-M^2\approx {\omega\over p}\left[ {\mathbf p}_\perp^2+M^2_{\rm eff}\right]\;{\rm with\ } 
M^2_{\rm eff}\equiv M^2 + {{Q^2}\over{\omega^2}}p_0 r_0\; .
\end{equation}
This virtuality can be used to write down the expression of the more
intuitive ``photon formation time'' $t_F$. Using the uncertainty principle,
we have indeed:
\begin{equation}
t_F^{-1}\sim \Delta E\sim {{R^2-M^2}\over{2r_0}}\sim {\omega\over{p_0 r_0}}
\left[ {\mathbf p}_\perp^2+M^2_{\rm eff}\right] \sim {{\omega M^2_{\rm eff}}\over{p_0 r_0}}
\end{equation}
We therefore arrive at an important observation: the collinear
enhancement is due to very small quark virtualities, which corresponds
to very long photon formation times. One can easily check that the
photon formation time increases if the photon becomes soft, or if the
invariant mass of the photon becomes very small.

There are a priori three other length scales that may also play a role in
this problem:
\begin{itemize}
\item The mean free path of the quarks in the plasma $\lambda_{\rm
    mfp}\sim (g^2T\ln(1/g))^{-1}$.
\item The range of the electric interactions $\lambda_{\rm el}\sim
  (gT)^{-1}$.
\item The range of magnetic interactions $\lambda_{\rm mag}\sim
  (g^2T)^{-1}$.
\end{itemize}
In the limit where $g\ll 1$, those scales satisfy $\lambda_{\rm el}\ll
\lambda_{\rm mfp} \ll \lambda_{\rm mag}$.

From there, one can check that the condition for having higher order
diagrams at the same order in $g$ as the bremsstrahlung is
$\lambda_{\rm mfp} \le t_F$, and that the relevant topologies
correspond to multiple scatterings of the quark in the medium
\cite{Gelis11,3-loop}. In other words, if producing the photon lasts
more than the typical time between two scatterings of the quarks, then
multiple scattering diagrams are also important.  This effect has
already been studied in a slightly different context, where a very
fast fermion is going through some cold medium. In that case, it is
known as the ``Landau-Pomeranchuk-Migdal'' effect. This effect reduces
the rate of radiative energy loss in the low energy end of the
spectrum. For photon production by a plasma, a preliminary (but
incomplete) study indicates that the LPM effect reduces the photon
rate both in the low end and in the high end of the spectrum
\cite{3-loop}.

\section{Nature of the relevant diagrams}
The last question we have to address is the nature of the multiple
scattering topologies that are dominant, and the short answer is that
it depends on the range of the interactions.

If the relevant interactions are short ranged, like the Debye-screened
electric interactions for which $\lambda_{\rm el}\ll \lambda_{\rm
  mfp}$, then on can check that only independent scatterings can
occur, as illustrated on the following figure:
\setbox1=\hbox to \textwidth{\hfill\resizebox*{!}{20mm}{\includegraphics{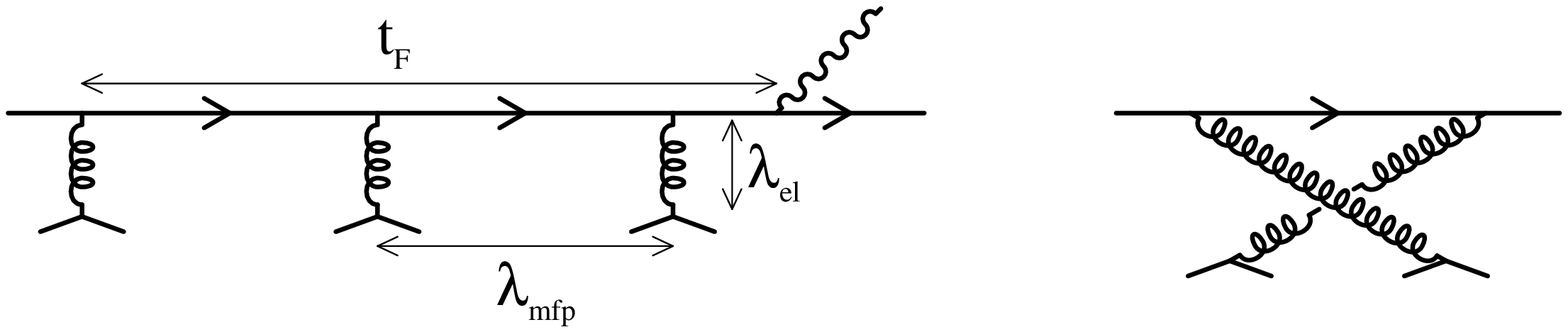}}\hfill}
\begin{displaymath}
\raise -11mm\box1
\end{displaymath}
In that case, only ladder topologies are important for the calculation
of the photon self-energy (together with the appropriate modification
of the quark propagators, in order to preserve gauge invariance).
Indeed, configurations like the diagram on the right in the previous figure
require an interaction range at least comparable to the mean free
path.

The situation can however become much more complicated if the
interactions are long ranged, like for static magnetic interactions,
since for them $\lambda_{\rm mfp}\ll \lambda_{\rm mag}$. In that case,
the above argument does not apply to restrict the set of relevant
topologies: successive scatterings are not independent from one
another and any topology can a priori contribute, unless some
unexpected cancellations occur. Those cancellations occur when the
process under study can only happen with hard scatterings, which means
that the relevant mean free path is not $\sim (g^2 T\ln(1/g))^{-1}$
(this is the average distance between two soft scatterings) but rather
$(g^4 T)^{-1}$ (which is the average distance between two hard
scatterings). When the relevant mean free path if larger than all the
interaction scales, then only ladder diagrams contribute.

Even if no extra cancellation occurs, a naive power counting seems to
indicate that sensitivity to the magnetic mass is suppressed by
$1/\ln(1/g)$ compared to the sensitivity on the mean free path.
Therefore, even in the worst scenario, one could calculate the photon
production rate at leading logarithmic accuracy by only summing ladder
diagrams. If no cancellation occurs, then terms beyond the leading
logarithm are truly non-perturbative.

\section{Conclusions}
The problem of collinear singularities in the photon rate by a plasma
is closely related to the interplay between the various distance
scales in the problem. Higher order topologies become important if the
photon formation time is larger that the quark mean free path, and the
nature and complexity of those topologies is controlled by the range
of the interactions compared to the mean free path.  Two questions can
probably be addressed analytically: the production of massive hard
photons is dominated by a single scattering if the invariant mass of
the lepton pair is large enough and one can probably perform the
resummation of ladder topologies, which can at least give the leading
log rate.

\noindent {\bf Acknowledgments:} My work is supported by DOE under grant DE-AC02-98CH10886.

\bibliographystyle{unsrt}

\end{document}